\documentclass[fonts]{icst}

\usepackage{moreverb}

\usepackage[breaklinks,colorlinks,bookmarksopen,bookmarksnumbered,linkcolor=ICSTblue,citecolor=ICSTblue,urlcolor=ICSTblue]{hyperref}
\usepackage{breakurl}
\usepackage{doi}

\usepackage{array}

\makeatletter

\providecommand{\tabularnewline}{\\}

\makeatother

\newcommand\BibTeX{{\rmfamily B\kern-.05em \textsc{i\kern-.025em b}\kern-.08em
T\kern-.1667em\lower.7ex\hbox{E}\kern-.125emX}}

\def\volumeyear{2014}

\begin{document}

\runningheads{M. Tsukada, J. Santa, S. Matsuura, T. Ernst, K. Fujikawa}{On the Experimental Evaluation of Vehicular Networks: Issues, Requirements and Methodology Applied to a Real Use Case}

\title{On the Experimental Evaluation of Vehicular Networks: Issues, Requirements and Methodology Applied to a Real Use Case}

\author{
Manabu Tsukada\affil{1}\affil{2}\fnoteref{1}, 
Jos\'{e} Santa\affil{3}\affil{4}, 
Satoshi Matsuura\affil{5}, 
Thierry Ernst\affil{6}, 
Kazutoshi Fujikawa\affil{7}
}

\address{
\affilnum{1} INRIA Paris - Rocquencourt, Domaine de Voluceau Rocquencourt - B.P. 105 78153 Le Chesnay Cedex, France\\
\affilnum{2} The University of Tokyo, 1-1-1, Yayoi, Bunkyo-ku, Tokyo, 113-8656 Japan\\
\affilnum{3} University Centre of Defence at the Spanish Air Force Academy, MDE-UPCT , Murcia, Spain\\
\affilnum{4} University of Murcia, Campus de Espinardo, 30100 Murcia, Spain\\
\affilnum{5} Tokyo Institute of Technology, 2-12-1, Ookayama, Meguro-ku, Tokyo, 152-8850, Japan\\
\affilnum{6} Centre de Robotique, MINES ParisTech, Paris, France\\
\affilnum{7} Nara Institute of Science and Technology, Nara, Japan
}

\abstract{
One of the most challenging fields in vehicular communications has been the experimental assessment of protocols and novel technologies. Researchers usually tend to simulate vehicular scenarios and/or partially validate new contributions in the area by using constrained testbeds and carrying out minor tests. In this line, the present work reviews the issues that pioneers in the area of vehicular communications and, in general, in telematics, have to deal with if they want to perform a good evaluation campaign by real testing. The key needs for a good experimental evaluation is the use of proper software tools for gathering testing data, post-processing and generating relevant figures of merit and, finally, properly showing the most important results. For this reason, a key contribution of this paper is the presentation of an evaluation environment called AnaVANET, which covers the previous needs. By using this tool and presenting a reference case of study, a generic testing methodology is described and applied. This way, the usage of the IPv6 protocol over a vehicle-to-vehicle routing protocol, and supporting IETF-based network mobility, is tested at the same time the main features of the AnaVANET system are presented. This work contributes in laying the foundations for a proper experimental evaluation of vehicular networks and will be useful for many researchers in the area.
}


\fnotetext[1]{Corresponding author.  Email: \email{tsukada@hongo.wide.ad.jp}}

\maketitle

\section{Introduction}

Intelligent Transportation Systems (ITS) are systems deployed to optimize the road traffic and realize safe, efficient and comfortable human mobility. There are a number of research fields in ITS but cooperative ITS and vehicular communications have received an especial attention during the last years. Within this area various technologies are considered, such as wireless communications, network management, communication security, navigation, etc.
In cooperative ITS, multiple entities share information and tasks to achieve common objectives. Thus, data exchange exists among vehicles, roadside infrastructure, traffic control centers, road users, road authorities and road operators, to support drivers, pedestrians, road authorities and operators in different areas of safety, traffic efficiency and infotainment. The European Commission (EC), for instance, published the action plan \cite{EC-COM-886:Action-plan} in Europe followed by ITS standardization mandate \cite{EC-M/453}, to speed up the adoption of these systems in the European Union, but there are a number of initiatives worldwide to encourage the research and development in ITS, mainly from the US Department of Transport and the Japan Ministry of Land, Infrastructure, Transport and Tourism. 

There are few barriers in the global road network among countries, and vehicles easily cross country borders, especially in Europe. Thus there is a huge necessity that cooperative ITS relies on the same architecture, protocols and technologies. As such, standardization organizations are developing cooperative ITS standards.
The International Organization for Standardization (ISO) Technical Committee 204 Working Group 16 (TC204 WG16) (also known as Communications Architecture for Land Mobile (CALM)) is in charge of standardizing a communication architecture for cooperative ITS. TC204 WG16 is specially working on a communication architecture supporting all types of access media and applications. In Europe, the European Telecommunications Standards Institute (ETSI) TC ITS is working on building blocks of the same architecture in harmonization with ISO TC204 WG16. In 2010, both ISO TC204 WG16 and ETSI TC ITS defined the ITS Station reference architecture \cite{ISO-21217-CALM-Arch, ETSI-EN-302-665-Arch}.
%
%

In cooperative ITS and, in general, in vehicular networks, there are two main communication paradigms, vehicle to vehicle (V2V) and vehicle to infrastructure (V2I), depending on whether the communication is performed directly between vehicles or using nodes locally or remotely installed on the road infrastructure.
When the V2V paradigm is considered, the research field is commonly called Vehicular Ad-hoc Networks, or VANET, as an especial case of Mobile Ad-hoc Networks (MANET) where nodes are vehicles. Although there are a lot of works related to VANET applications and basic research at physical, MAC and network layers, there is a significant lack of real evaluation analysis in this field, due to cost and effort implications. 
Many VANET solutions and protocols could be considered as non-practical designs if they were tested over real scenarios, as it has been proved in MANET~\cite{Tschudin03}. Performance of VANET protocols based on a pure broadcast
approach can be more or less expected in simple configurations, even if they are
not experimentally tested; but the number of issues concerning the real performance of multi-hop designs is much more tricky.  A similar problem can be found in V2I, which has received a great attention by the research community in the last years, due to the idea that V2I technologies and services will find a place in the market before V2V approaches. Nevertheless, a number of experimentation works and supporting tools should be improved in the short term, in order to give real evidences to car manufacturers and road operators of the benefits of vehicular communications.

\begin{figure}[htbp]
  \begin{center}
    \href{http://anavanet.net/demo-vienna/?analysis=1264516500}{
    \includegraphics[width=\linewidth,clip]{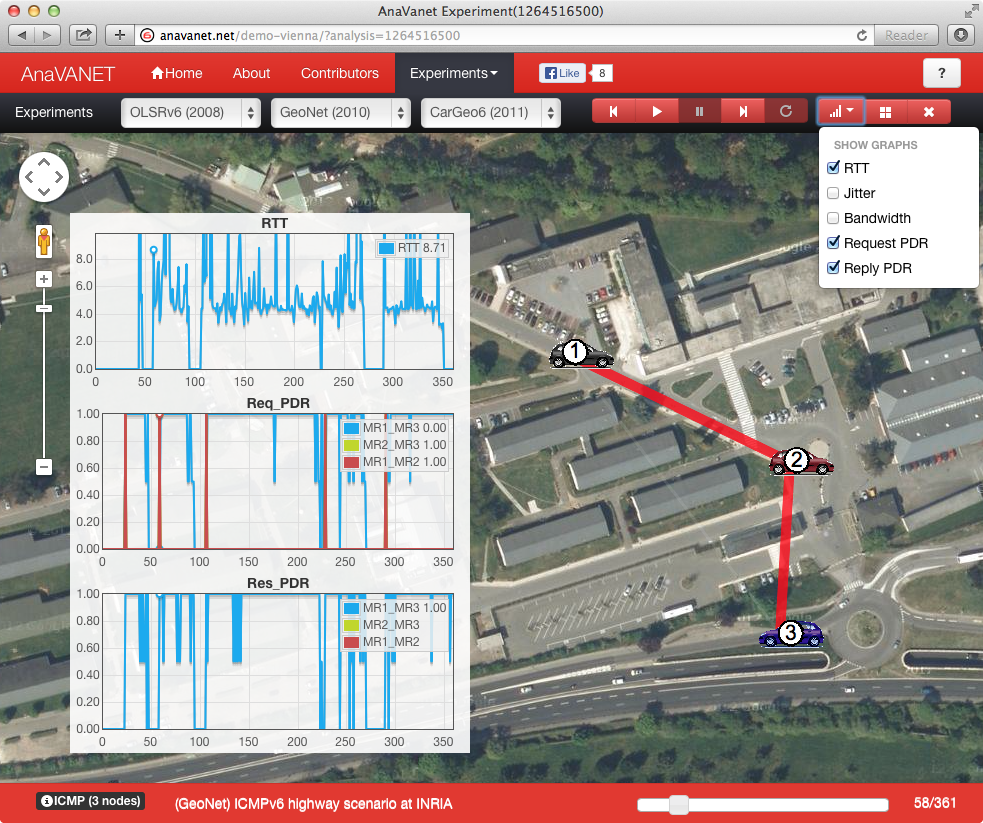}} 
  \caption{Screenshot of AnaVANET viewer}
  \label{fig:screenshot}
  \end{center}
\end{figure}

Conventional network measurement tools (e.g. \textit{iperf}, \textit{ping} or \textit{traceroute}) assume fixed networks and assess network performances in an end-to-end
basis. However, under dynamic network conditions such as in the vehicular networks case, it is difficult to analyze in detail the operation of networks by using solely these tools, because vehicles are always changing their location and the performance of wireless channels fluctuates. In order to solve
these issues, we have developed a packet analysis and visualization tool called
AnaVANET\footnote{\textit{\href{http://anavanet.net/}{http://anavanet.net/}}}, which considers the peculiarities of the vehicular environment for providing an exhaustive evaluation software for outdoor scenarios (Figure~\ref{fig:screenshot} includes a preliminary screenshot of the visualization). 
Both V2V and V2I networks can be efficiently analyzed, thanks to the integrated features for collecting results, post-processing data, generate graphical figures of merit and, finally, publish the results in a dedicated web site (if desired). 
All tests and results are later available in the form of an animated webpage where both researchers and the general public can access the evaluations. 
AnaVANET has been successfully exploited for the moment in experimental evaluation campaigns in the GeoNet~\cite{GeoNet-D.7} and ITSSv6~\cite{ITSSv6-D4.2} projects.

In this paper, apart from presenting our tool for assessing the performance of vehicular networks, we analyze in detail the problem of real testing in V2V and V2I, identifying the main issues, requirements, and proposing a general methodology useful for further works in the area. To sum up, the rest of the paper is organized as follows.  
Section~\ref{lbl:background} introduces the readers in network layer protocols for vehicular networks. Section~\ref{lbl:experimental} reviews related works in the area of testing vehicular networks. Then, the issues and requirements for evaluating vehicular networks are listed in Section~\ref{lbl:issues}. 
The evaluation methodology desired in this frame is described in Section~\ref{lbl:methodology} and, as a result of our analysis, the design and implementation of the AnaVANET evaluation tool is detailed in Section~\ref{lbl:anavanet}, together with a reference evaluation of a network testbed using the tool in Section~\ref{lbl:evaluation}. As a result of this evaluation, the functionalities provided by AnaVANET are analyzed in Section~\ref{lbl:qualitative} according to the previously identified needs. Finally, Section~\ref{lbl:conclusion} concludes the paper summarizing the main results and addressing future works. 




\section{Network protocols in vehicular networks}
\label{lbl:background}

\begin{figure*}[htbp]
  \begin{center}
    \includegraphics[width=0.9\linewidth]{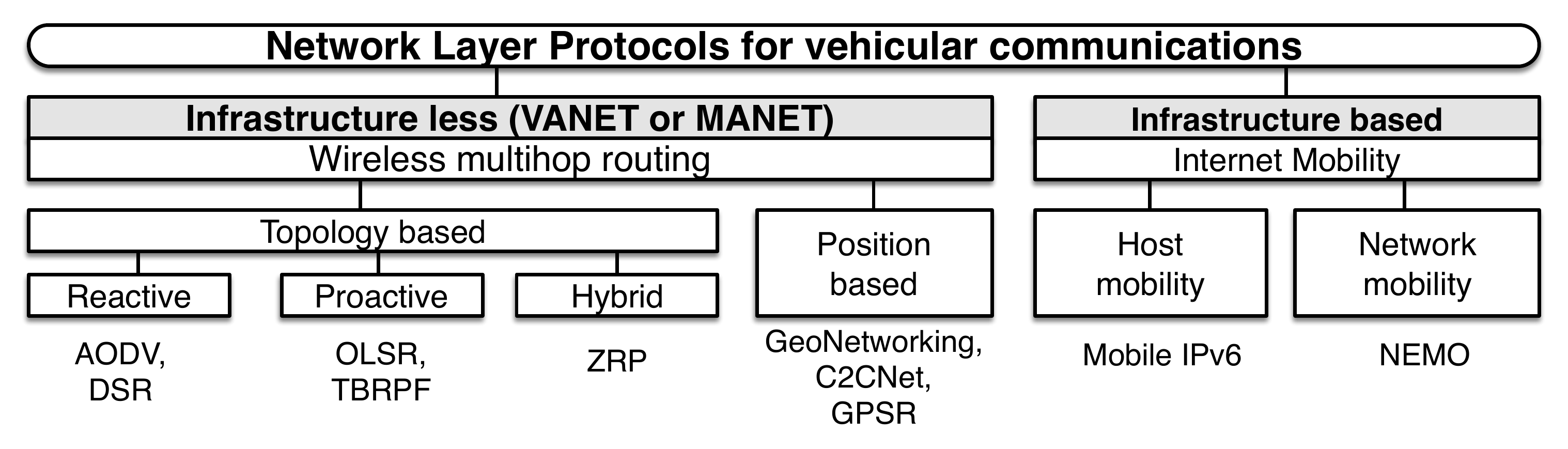}
  \caption{Network protocols in vehicular networks}
  \label{fig:routing-protocols}
  \end{center}
\end{figure*}

Network protocols in vehicular networks can be classified in
infrastructure-less scenarios, i.e. V2V,  and infrastructure-based scenarios, i.e. V2I, as showed in Figure~\ref{fig:routing-protocols}. 

The infrastructure-less scenario is well-known in the research areas of VANET and MANET. These approaches are designed to enable wireless
communications in dynamic topologies without any infrastructure. Routing
protocols here are further classified as \textit{topology-based} and
\textit{position-based} routing protocols. Upon the appearance of vehicular
communications, a second class of infrastructure-less protocols added to the
list: VANETs. Most of the VANET solutions are based on geographical routing,
thus based on the node's position. 

Topology-based protocols were divided into two main branches by the IETF MANET
working group: \textit{reactive}, where nodes periodically exchange messages to
create routes, and \textit{proactive}, in which control messages are exchanged on
demand when it is necessary to reach a particular node. Generally, proactive
protocols have the advantage of starting communication rapidly by making the
routing table ahead, however, this makes battery life shorter due to frequent
signaling. If the topology is highly dynamic and the data traffic is frequent, a
proactive protocol could be better. Reactive protocols, on the contrary, keep
the battery life longer by reducing signaling messages when there is no data to
transmit. The \textit{hybrid} protocols that take the advantage of both
proactive and reactive protocols by maintaining routes to near
neighbors regularly and searching the destination in long distance on demand.

Some routing protocols are specified by the IETF MANET working group~\cite{manetwg}.
Both IPv4 and IPv6 are supported in the working group. Ad hoc On-Demand Distance
Vector Routing (AODV)~\cite{rfc3561} and Dynamic Source Routing Protocol (DSR)
\cite{rfc4728} are specified as reactive routing protocols. And Optimized Link
State Routing (OLSR)~\cite{rfc3626} and Topology Dissemination Based on
Reverse-Path Forwarding (TBRPF)~\cite{rfc3684} are specified as proactive
routing protocols. As an example of a hybrid MANET protocol, Zone Routing Protocol
(ZRP)~\cite{ZRP} is proposed. 

VANETs are a particular case of MANETs, and are
not restricted by the battery of the communication nodes and are also characterized by the high speed of nodes, the availability of GPS information, and a regular distribution and foreseeable movements. First, vehicles have a larger battery than mobile terminals or sensor devices, which is also charged when the engine is
running. Second, the speed of vehicles is also higher than common portable
terminals, and relative speeds can reach 300 Km/h; hence, the duration of the
routing entries is extremely short. Third, a GPS device and digital map can be assumed in many
cases, whose information improves the network performance in some proposals.

Unlike topology based routing, position based routing does not need to maintain
part of the network structure in order to forward packets towards the
destination node. When the routing is based on the position, nodes forward the
packets with the aim of reaching the nodes within a geographical location. Thus,
position based routing can eliminate the problem that appears in topology based
protocols when routes become quickly unavailable in high mobility scenarios. In
Greedy Perimeter Stateless Routing (GPSR)~\cite{Karp2000}, for instance, 
intermediate nodes make a decision based on the destination position and
neighbor positions. The Car-to-Car Communication Consortium (C2CC) also
specified the C2CNet protocol, which was later enhanced by the GeoNet project to
support IPv6. Within the ITS standardization domain,
GeoNetworking~\cite{ETSI-TS-102-636-4-1-Media} is being completed by ETSI at the
moment, integrating several geo-aware strategies to better route packets in
vehicular networks.

On the other side, infrastructure-based protocols have been focused on the
global connectivity of nodes to the Internet. Mobile IPv6~\cite{rfc6275} solved
the mobility problem for mobile hosts and, later, Network Mobility Basic support
(NEMO)~\cite{rfc3963} provided a solution for the mobility of a whole network
(e.g. a vehicle or bus), which has been recommended by the ISO TC204 WG16 to
achieve Internet mobility for vehicles. NEMO maintains a bi-directional tunnel between the router in the vehicle, known as the \textit{mobile router (MR)}, and a server in the fixed infrastructure, known as the \textit{home agent (HA)}, in order to provide a unchanged network prefix called \textit{mobile network prefix (MNP)} to the in-vehicle network. All the in-vehicle nodes called \textit{mobile network nodes (MNN)} maintain a permanent address derived from the MNP even when MR changes the point of attachment to the Internet during the movement (i.e. handover).



\section{Experimental evaluation of VANET approaches in the literature}
\label{lbl:experimental}

Because of equipment cost, logistic issues and, in general, the necessary effort, literature in experimental evaluation of vehicular network architectures is limited. However, these works are of key importance for the ITS community. Up to now, there are several works dealing
with this issue, although most of them are still focused on studying the
operation of WiFi, DSRC (Dedicated Short Range Communications) or IEEE 802.11p technologies in the vehicular field.

Communication between a vehicle and a static terminal is important for some
ITS services. In~\cite{Wewetzer2007} a communication scenario considering a
static terminal and a moving vehicle is studied in detail. Among all metrics
considered in this work, the transmission power is the more original one,
determining the maximum communication range. The type of data traffic used to
test the performance of the communication channel is also of interest. Most
VANET designs use UDP packets, due to poor TCP performance over wireless
channels~\cite{Hui05, festag4fbc}. The evaluations performed in~\cite{shagdar:hal-00702923} with IEEE 802.11p  reveal that the packet delivery ratio achieved by this technology is highly dependent on the distance between sender and receiver. These results are also confirmed in~\cite{Lin12}, where it is also concluded that the vehicle speed does not imply a noticeable performance degradation of the communication. A similar evaluation is performed in~\cite{Gozalvez12}, but this time carrying out a great testing campaign in a city. 
 

When V2V scenarios are considered, most of the previous works only consider two terminals in performance tests, what is not too representative when multi-hop schemes are evaluated. In~\cite{1390578}, the applicability of 802.11b in V2V communications is evaluated over urban and highway scenarios, and it is
demonstrated that a direct line of sight is one of the most important issues in the network performance.
Two works evaluate a multi-hop VANET over real conditions, using three \cite{Jerbi2007c} and even six vehicles
\cite{Jerbi2008}. These papers offer a wide study about a real VANET set-up, and
the last one includes an interesting analysis describing the impact of the
number of hops on the final performance.
Nonetheless, static routes are used in that work, presenting a non-realistic vehicular network. Our prior work~\cite{Tsukada2010a}, by contrast, considered a real and standardized ad-hoc routing protocol to dynamically modify communication paths. The hardware testbed presented is also suited for future ITS research, with a flexible in-vehicle and inter-vehicle IPv6 network based on mobile routers.

To the best of our knowledge, there exists a few works dealing with the evaluation of IPv6-based communications at network level in vehicular communications, and some of them are within our research line~\cite{Santa13b,Santa13a}. However, our prior evaluations are only focused on IPv6 network mobility. In this work, the operation of NEMO over a V2V protocol is evaluated, using an implementation of GeoNetworking. This way, an integrated V2V and V2I approach is considered for providing an integral vehicular connectivity using IETF and ETSI standardized protocols. The novelty of this work is twofold, since not only this routing approach is experimentally analyzed, but also an evaluation tool especially designed for vehicular networks is used. As far as the authors know, no specific tools for assessing the performance of vehicular networks have been developed or used in previous research works.


\section{VANET evaluation: issues and requirements}
\label{lbl:issues}

\subsection{Issues}

As said above, the experimental evaluations carried out in vehicular networks are mostly based on single-hop studies. In the case of multi-hop experiments, a static route configuration is often employed, but dynamic routing presents a more realistic view in vehicular communications. 

Using multi-hop and dynamic routing strategies presents a challenge in the evaluation of vehicular networks.
Common end-to-end evaluation tools such as \textit{ping6} and \textit{iperf} are useless to track the effect of route change, because they are unaware of the path taken during a communication test. An additional lack of these tools is the possibility to measure the performance of hop-by-hop links, since the study is carried out end-to-end. Also, geographical and external factors such as nodes position, distance between nodes or obstacles are not linked with network performance figures of merit.
Therefore, the performance comparison of various dynamic routing
protocols is essentially missing. 

\subsection{Requirements}
\label{lbl:requirements}

With the aim of summarizing these main requirements when evaluating multi-hop vehicular networks, the next needs are found essential by the software tools used in experimental campaigns for evaluating both V2V and V2I:

\subsubsection{Path detection} 

The topology and communication path of a vehicular network changes frequently with dynamic routing as
vehicles move. Thus, the tool should take note of the communication path used in every moment.

\subsubsection{Communication performance in links} 

The communication performance between ends is the sum of the links on the way
between them. Once the communication path is tracked, the tool should measure
the performance link by link as well as end-to-end. 

\subsubsection{Geographical awareness} 

The network performance in a link depends on various geographical factors. For
example, the distance between the nodes affects the packet loss probability of
the link; the movement speed and the direction are also important factors for
the packet loss in the link; and the existence of obstacles between the nodes
may screen the wireless radio propagation. Thus, the evaluation tool should take
the above geographical factors into account. 

\subsubsection{Intuitive visualization}

It is important to visualize the geographical factors such as node movement
(speed, direction), distance and signal obstacles in order to analyze which of them
affect the network performance. For intuitive visualization, performance figures
of merit and environmental information should be shown together in a
synchronized way. Moreover, the spatio-temporal data series should be available
in post process to play them at different speeds, stop when desired, or replayed
freely as he or she wants. 

\subsubsection{Independence from network protocols}

As shown in Section~\ref{lbl:background}, there are many network layer protocols
in the literature for vehicular scenarios, both infrastructure-less and
infrastructure based. The evaluation tool should be independent from the network
protocols employed in the target vehicular network. This includes that the tool
does not require changes to adapt to neither specific protocols nor special
message or data transported. 

\subsubsection{Independent from devices} 

Depending on the experiment, the configuration of the used devices may differ in
both vehicle and infrastructure sides. The devices include the antenna, wireless
chipset, CPU, memory, GPS and so on. The tools should not rely on any of the
specific devices functionalities. Most favorably, the same software and 
settings for an experimental test should work on multiple devices. 

\subsubsection{Adaptation to various scenarios}

There are a number of possible networking scenarios in vehicular communications,
such as using parked vehicles, slow speeds with surrounding buildings in a urban
situation, vehicles moving at higher speeds in a highway, overtaking, vehicles
crossing in a two-way road, different topological locations of the ends in a V2I
setting, etc. The software evaluation tool should accommodate to all of these
scenarios.   

\subsubsection{Easiness for data collection}

In order to compare the network performance obtained when using different
network protocols, a lot of experiments could be needed. This may require
installing data collector software on many devices, depending on the scenario.
Thus, the easiness of the installation of these software modules is very
important. Of course, the most favorable case is to employ common software in
all of them, such as \textit{tcpdump} or \textit{cat}. 
 
\subsubsection{Flexible experimental data format}

The experimental data should be stored in a well-organized way. Therefore, the data
format needs to be flexible for future extension. For example, the user of the
system could require adding new attributes to the data format of evaluation
results. We must consider flexible data formats in order not to impact the
process of adding new attributes.


\section{Evaluation methodology}
\label{lbl:methodology}

As it is later described, the evaluation tool presented in this work (AnaVANET) copes with the previous requirements, but first it is important to identify a generic testing methodology that allow a researcher to success in a testing campaign with a vehicular network.

\begin{figure*}[bhpt]
   \begin{center}
       \includegraphics[width=0.9\linewidth]{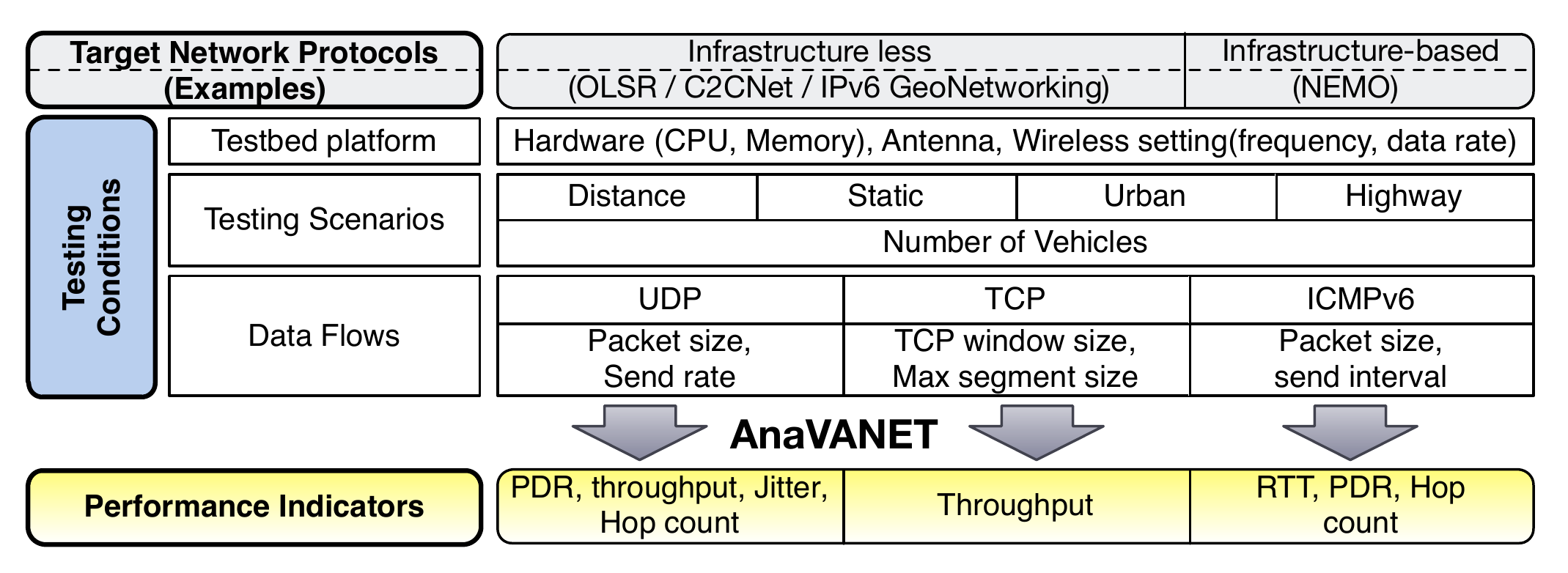}
  \caption{Evaluation methodology}
  \label{fig:methodology}
  \end{center}
\end{figure*}

In general, the evaluation goals in computer networks are to analyze which \textit{testing conditions} affect which \textit{data flows or network protocols}. For achieving this end it is necessary to design a proper evaluation methodology. Within it, we should consider the tendency of results by repeating tests with the same settings or varying parameters under study, such as the network protocol, the mobility of nodes or the data volume. A proper evaluation tool, such as the later presented AnaVANET, should support the overall analysis.  This section considers both the testing conditions and the possible routing protocols to consider in vehicular networks, as it is summarized in Figure~\ref{fig:methodology}, by introducing the concept and presenting our real use case for testing the performance of NEMO over IPv6 GeoNetworking. 

\subsection{Testing conditions}

\subsubsection{Testbed platform} 

The testbed used for the evaluation of network architecture should be carefully chosen to implement most relevant nodes in real software and hardware. In vehicular communications, this is extremely important, since a good deployment could be needed in case of testing V2V multi-hop networks.

\begin{figure}[htbp]
   \begin{center}
    \includegraphics[width=1\linewidth]{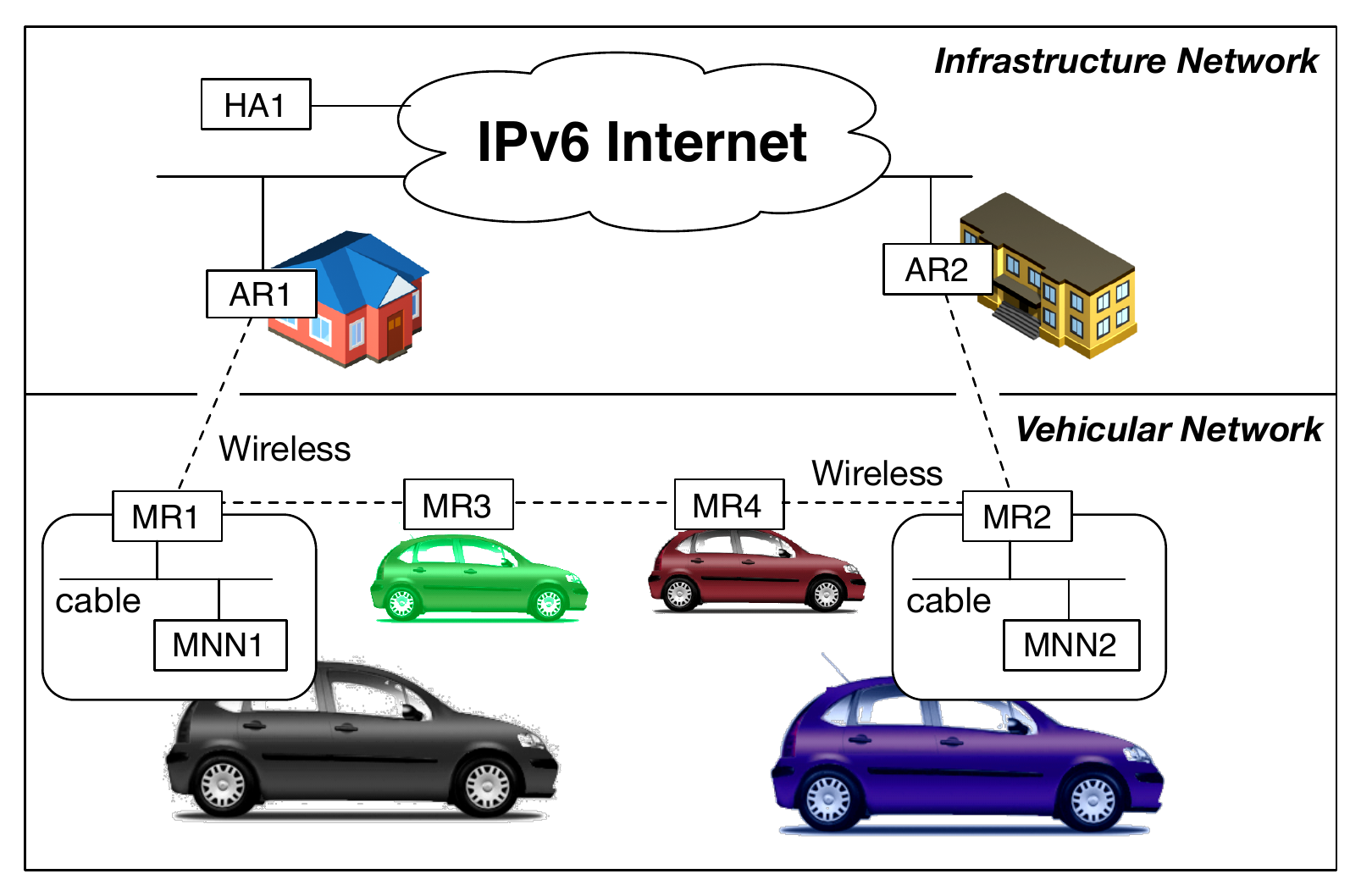}
      \caption{Reference network configuration}
      \label{fig:network-configuration}
  \end{center}
\end{figure}

In our particular case, the testbed comprises a set of four vehicles and two roadside stations, as illustrated in Figure~\ref{fig:network-configuration}. Each vehicle is equipped with a mobile
router (MR), with at least two interfaces: an Ethernet link to connect mobile
network nodes (MNNs) within the in-vehicle network, and a wireless adapter in ad-hoc mode
used for both V2V and V2I communications. On the roadside, access routers (ARs) are fixed on the top of a building or any other elevated point near the road. Each one provides two interfaces: an Ethernet link for a wired Internet access, and a wireless adapter in
ad-hoc mode to connect with vehicles in the surroundings. At a backend point on the Internet, a home agent (HA) is installed to support Internet mobility of MRs by using NEMO. 

Among the various testbed conditions, the hardware specifications (CPU, memory, etc),
antenna and wireless settings are important factors for the evaluation, since they will highly affect the results. In our case, MRs are Alix3d3 embedded boxes provided with a Linux 2.6.29.6 kernel. Each MR has a mini-pci wireless card Atheros AR5414 802.11 a/b/g Rev 0, and
an antenna 2.4GHz 9dBi indoor OMNI RP-SMA6 is used. The frequency used has been 2.422Ghz and the data rate has been fixed to 6 Mbits/s. 


\subsubsection{Testing scenarios}

\begin{figure}[htbp]
   \begin{center}
    \includegraphics[width=1\linewidth]{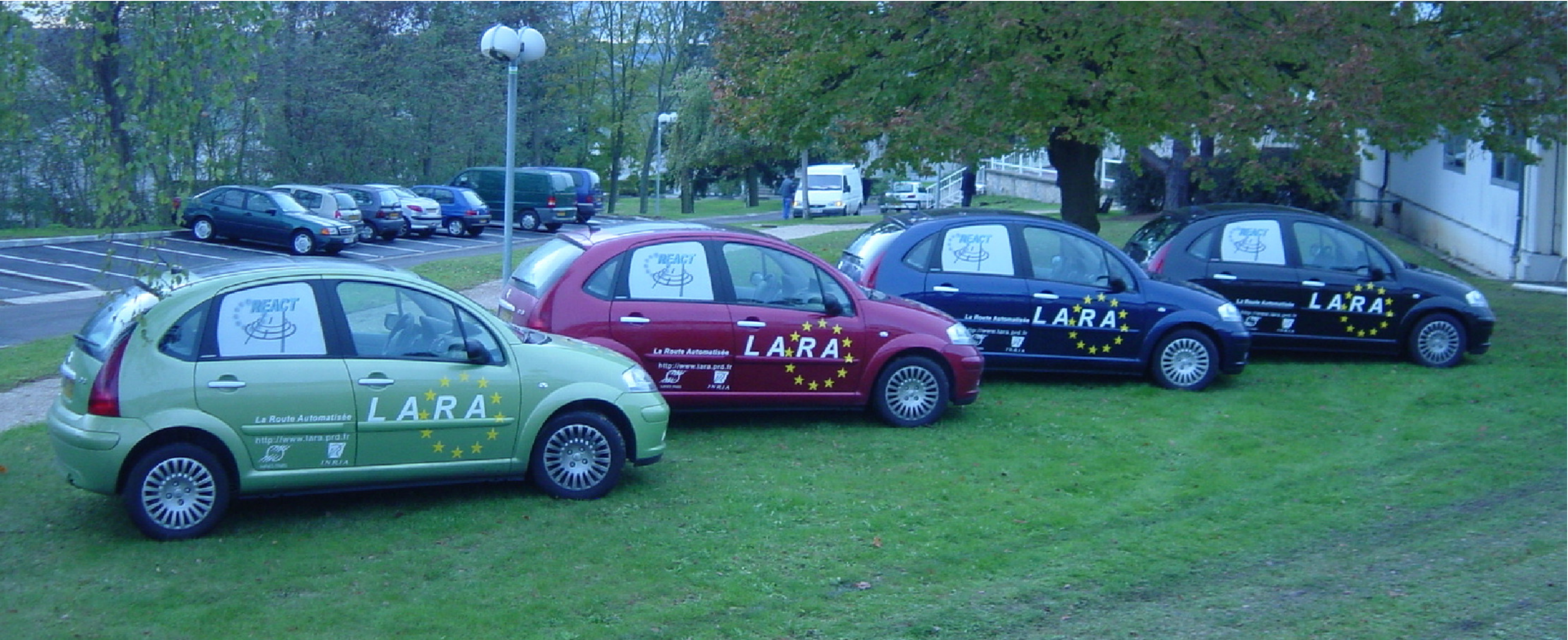}
      \caption{Testing vehicles}
      \label{fig:Prototype-vehicles-used}
  \end{center}
\end{figure}

Fixing the evaluation scenarios beforehand is essential in the planning of a testing campaign. In general, the main factors that determine the possible scenarios are:

\begin{description}

\item [\bf Mobility] Vehicle mobility is a key issue to cope with realistic
      vehicular network conditions. This way, we can consider not only static
      scenarios, to test the network operation in a controlled way, but also
      dynamic scenarios under common speed situations. Of course, field
      operational tests should be conducted to confirm the expected results,
      taking into account the proper handling of mobility, i.e. Doppler shifting,
      fast fading, etc.

\item [\bf Location] Urban and interurban environments affect communication
      performance in a different way, because the signal propagation can be
      interfered by buildings (among other elements), and the line of sight
      between vehicles is not always possible. Two environments are considered
      in our tests: a semi-urban one located at INRIA-Rocquencourt, which
      contains a set of small buildings surrounded by streets, and a highway
      stretch, the A-12 one, near INRIA-Rocquencourt. 

\item [\bf Number of vehicles] The number of hops between the source and the
      destination vehicles affect the communication delay and the higher
      probability of packet looses, due to route changes or MAC transmission
      issues. Up to four conventional vehicles (Citro{\"e}n C3)  are considered in our case. This testing fleet is
      showed in Figure~\ref{fig:Prototype-vehicles-used}. 

\end{description}

\begin{figure}[htbp]
   \begin{center}
    \includegraphics[width=1\linewidth]{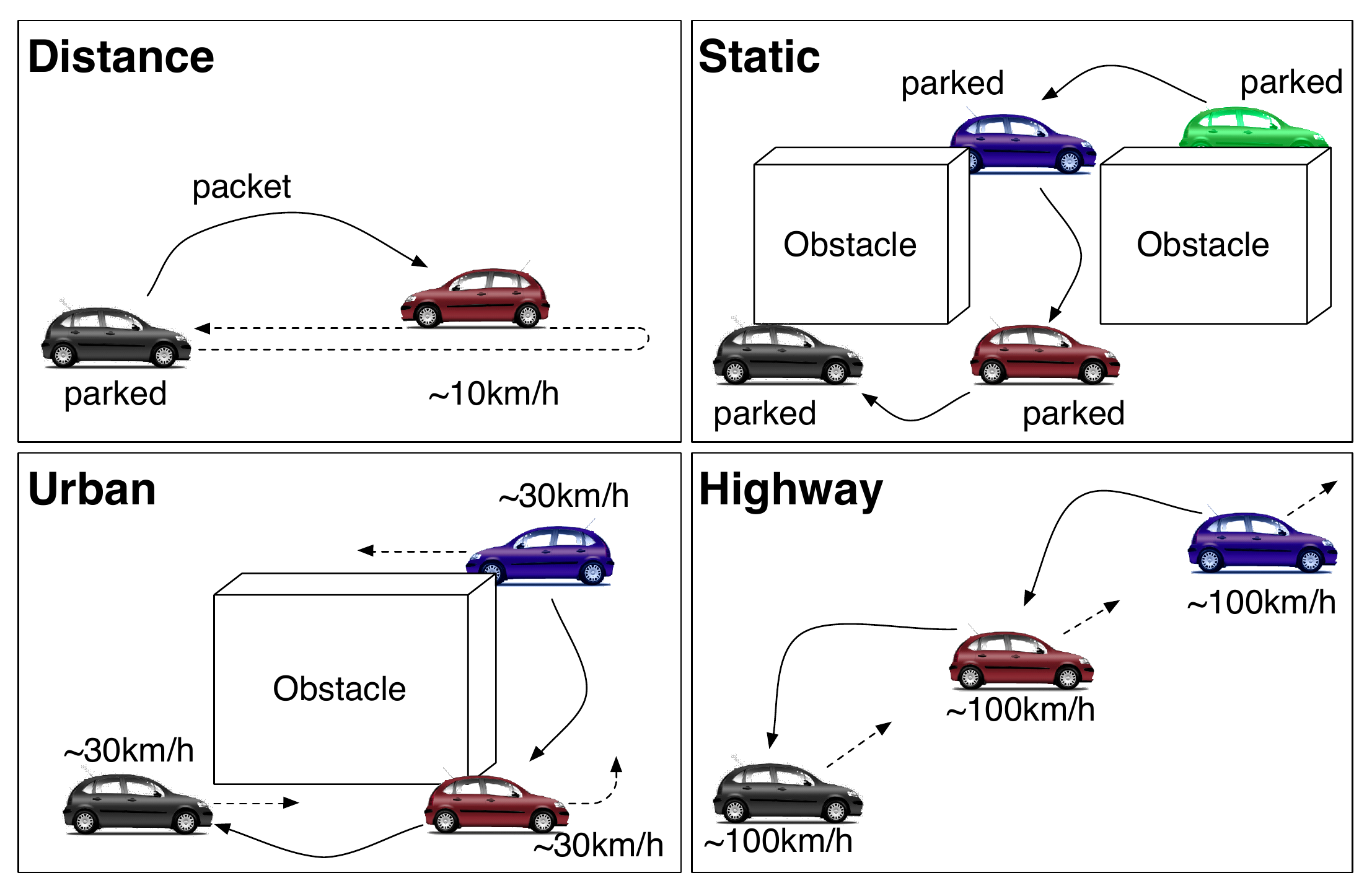}
      \caption{Proposal of movement scenarios}
      \label{fig:scenarios}
  \end{center}
\end{figure}

A set of possible testing scenarios when evaluating multi-hop vehicular networks is summarized in Figure~\ref{fig:scenarios}. These have been divided
into urban and highway. Mobility has been set to static, urban-like speed and
high speed. In our particular evaluation, these scenarios have been considered with our fleet of vehicles, with the aim of covering a wide range of communication conditions. The obstacles have been in our case a set of building blocks located at the Paris - Rocquencourt premises. The chosen highway has been the French
A13, near Versalles.

\subsection{Data flows and performance indicators}

A number or protocols and data flows can be set for evaluations, however, only
the most representative and more used in the literature should be considered to
study concrete performance indicators. For instance, in our case UDP, TCP and
ICMPv6 are used to measure the network performance between two communication
end-nodes (MNN to MNN) mounted within two vehicles:

\begin{description}

\item [\bf UDP] is a connection-less unidirectional transmission flow. The
      traffic is generated by \textit{iperf} in our case. It is considered that with UDP the performance indicators under consideration can be the packet delivery ratio, throughput and jitter.
      
\item [\bf TCP] is a connection-oriented bidirectional transmission flow. This
      traffic is also generated by \textit{iperf} in our case. The performance indicator under consideration here has been the maximum throughput.
      
\item [\bf ICMPv6] is a bi-directional transmission flow. The traffic is
      generated by \textit{ping6} in our case. The performance indicator under
      consideration can be the road trip delay time and packet deliveries.

\end{description}

The set of performance indicators most used in the literature are detailed next:

\begin{description}

\item [Round-Trip Time (RTT)] can be measured using ICMPv6, as in our case. A host on the source
      vehicle, or located at an infrastructure point, sends ICMPv6 echo request to a host on the destination
      vehicle, or located at an infrastructure point. The destination host replies with an ICMPv6 echo reply. The period between the time that the request is sent and the time that the reply is
      received can be obtained by using ping6.

\item [Throughput] can be measured using UDP or TCP. It can be measured with a traffic generator tool, such as the \textit{iperf} tool in our case. In UDP, iperf is executed in both the sender and the receiver nodes. The UDP
      packet transmission rate is set with a fixed rate and the sender is not able to see the
      result because the communication is unidirectional from the sender to the
      receiver. The throughput is shown on the receiver side. On the other hand,
      when using a TCP transmission, the sending rate is automatically adjusted with the TCP congestion control
      mechanism. The sending rate is adjusted depending on the acknowledgement
      messages received. The throughput appears in both the sender and receiver nodes.               

\item [Jitter] is a measure of the variability over time of the packet latency
      across a network. A network with a constant latency has a null jitter. In general, the jitter is expressed as an average of the deviation from
      the network mean latency, and can be calculated using the RTT, as in our case.

\item [Packet Delivery Ratio (PDR)] is the percentage of packets received by 
      the target node as compared with the number of packets sent by
      the source. iperf, for instance, shows this value at the receiver side when using TCP in an end-to-end manner, but AnaVANET is also able to calculate the PDR on each hop between the sender and destination nodes.

\end{description}


\section{System design and implementation of AnaVANET}
\label{lbl:anavanet}

\subsection{Overview of the software}

AnaVANET (initially standing for Analyzer of VANET) is an evaluation tool implemented in Java to assess the performance of vehicular networks. It takes as input the logs generated by the \textit{iperf}, \textit{tcpdump} and/or \textit{ping6}, together with navigation information in NMEA format, to compute the next performance metrics: network throughput, delay, jitter, hop count and list of intermediate nodes in the communication path, PDR end-to-end and hop-by-hop, speed, and instantaneous position. 

\begin{figure*}[htbp]
   \begin{center}
    \includegraphics[width=0.8\linewidth]{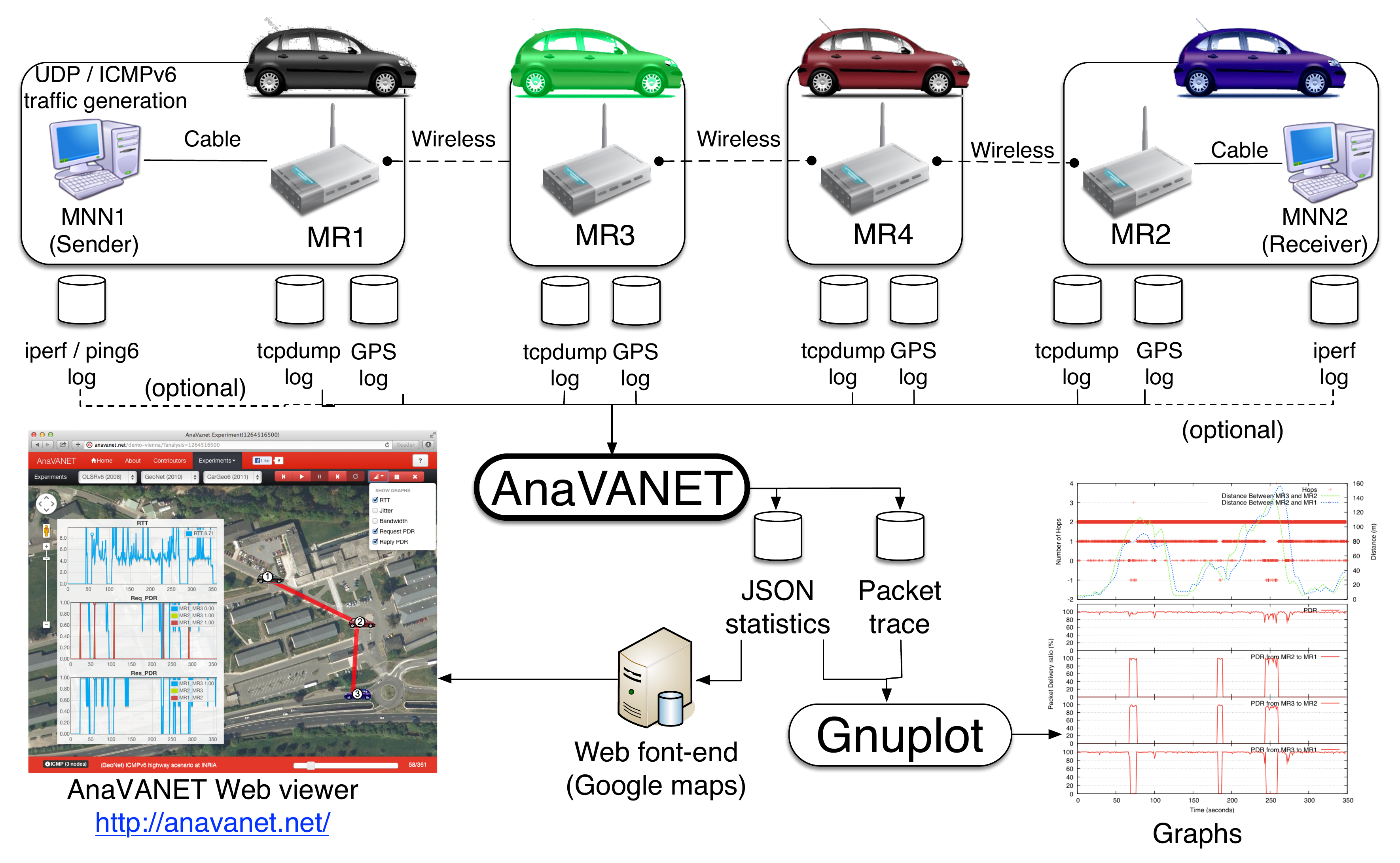}
  \caption{Overview of AnaVANET}
  \label{fig:anavanet}
  \end{center}
\end{figure*}

In this part of the work AnaVANET is put in the context of the evaluation scenario described in the previous section in Figure~\ref{fig:anavanet}, showing also the main inputs and outputs of the tool. The sender MNN (left most vehicle) is in charge of generating data traffic, and both the sender and the receiver (right most vehicle) MNNs record a high level log,
according to the application used to generate network traffic (\textit{iperf} and \textit{ping6} for the moment). All MRs record information about forwarded data packets by means of the \textit{tcpdump} tool, and log the vehicle position continuously. All this data is post-processed by the
AnaVANET core software and then analyzed. The tool traces all the data
packets transmitted from the sender node to detect
packet losses and calculate statistics for each link and end-to-end, and then merge
all these per-hop information with transport level statistics of the traffic
generator. As a result, AnaVANET outputs a JSON file with statistics on a
one-second basis (see Section\ref{lbl:data-format} for details), and a packet trace file with the path followed by each data packet. 

Once generated, performance metrics can be graphically showed through plots generated by \textit{gnuplot} and a website where all tests are available. The screenshot of the website is shown on the left bottom corner of Figure~\ref{fig:anavanet} (which is also enlarged in the previous Figure~\ref{fig:screenshot}). Accessing the website one can replay the tests on a map to see momentary figures of merit. 
Previous experiments can be chosen to monitor the main performance metrics at any time of the tests. Users can play and stop at any arbitrary point of the test with the
control buttons on the upper left part of the window. The player speed, one step forward
and one step backward are also implemented.
On the map, the position and
movement of the vehicle are depicted with the speed of each vehicle and the
distance between them. The transferred data size, bandwidth, packet loss rate,
RTT and jitter, for each link and end-to-end are displayed. The network
performance is visualized by the width of links and the colors used to
draw them.

%

%
%
%
%

\subsection{Data format of experimental results}
\label{lbl:data-format}

In this part, we describe the problems of the former AnaVANET data format~\cite{Tsukada2014}, which was based on XML, and we detail the recent changes to improve the flexibility of the results using the JSON format~\cite{rfc7159}.

There is a fundamental trade-off between flat data format and structured data
format. Flat data format is more flexible than structured one, because if
developers want to add a new attribute, they just put the attribute next to the other
attributes. On the other hand, in a structured format, developers
have to consider the layers and relationships to add a new attribute and they
sometimes cannot add the new attribute because of its structure. However, when a flat data format is used, developers have to revise and adjust their applications, since the relationships among
attributes can vary. In this line, a normalized way of calling the attributes is also important. If there is no
rule of normalization, developers have to handle differences of an attribute
name (e.g., temperature, Temperature, temp). 

AnaVANET was initially developed to analyze the real operation of VANETs. The initial data format used as output of an evaluation had some problems regarding its flat format and the dynamic columns available per each data record. Hence it took several hours to check the results after carrying out new experiments. To
solve these problems, we have designed a structured and normalized format,
considering the features of vehicular networks. Our format is extensible and independent from concrete experimental environments and visualization tools. We have also adapted the initial visualization tool with an internal converter module within the web application, and an additional command line tool has been implemented to process the output logs of AnaVANET in a text-based basis. The new data format and tools enable us to check the results in several seconds after carrying out experiments, considering that users could require a fast evaluation to continue with new experiments.

\begin{figure}[htbp]
   \begin{center}
       \includegraphics[width=1\linewidth]{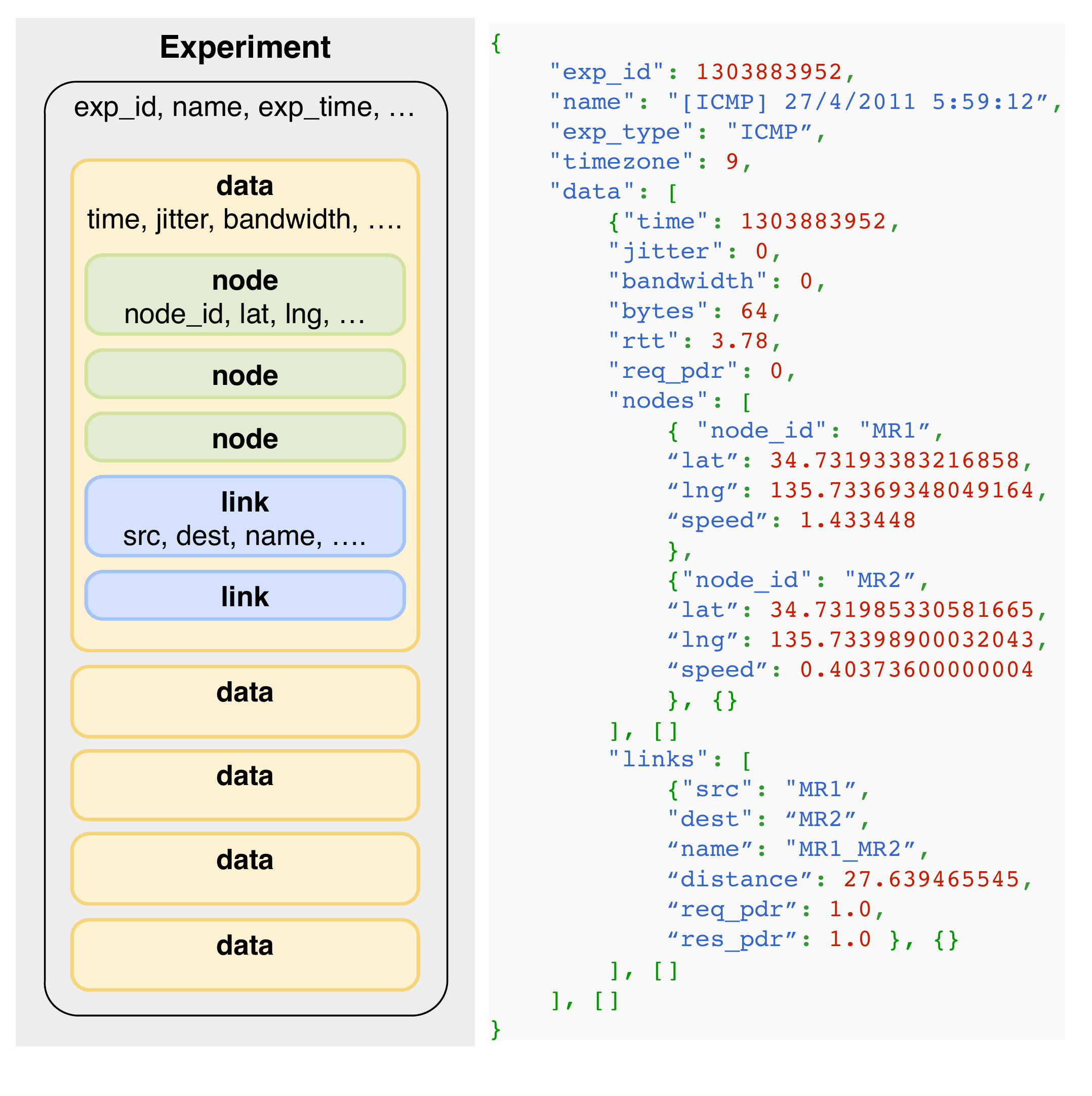}
      \caption{Three-layer model of the structured data format used by AnaVANET}
      \label{fig:data-format}
  \end{center}
\end{figure}

The new structured and normalized data format of AnaVANET considers three layers, as it is shown in Figure~\ref{fig:data-format}. AnaVANET summarizes data on each time slot following the next scheme. The top layer is the ``experiment'' layer. This layer mainly manages static attributes (e.g. ID of experiment or the name of
experiment). The second layer is the ``data'' layer. This layer manages results of an
experiment on each time slot. This layer has time, total packet delivery ratio
(PDR), total RTT and other attributes. The third layer is comprised by the ``node'' and ``link'' parts.
The node part manages each node's statuses, whereas the link part manages each link's
statuses. Link means a relationship between two nodes and especially represents
wireless link statuses. An experiment has a series of data and each data has
several nodes and links. We have also normalized the names of the attributes considered in each layer.

This data format based on time-series for saving node and link information is an abstracted representation that can be used to collect results from any kind of network. Moreover, by using this three-layer representation, the system can be easily adapted to future requirements. 

\section{Evaluation of NEMO over IPv6 GeoNetworking} 
\label{lbl:evaluation} 

Early versions of AnaVANET were designed for evaluating infrastructure less
network protocols, as used in our previous works for
analyzing OLSR in vehicular environments~\cite{Santa2009a} and later tests of IPv6 over C2CNet~\cite{Tsukada2010b} in the FP7 GeoNet project. The current version of AnaVANET can also analyze infrastructure-based network
protocols such as NEMO. 

In this section, we report a summary of the results collected in the evaluation
of NEMO over IPv6 GeoNetworking when a vehicle connects with a node located in
the Internet using two roadside units as access routers. The
\textit{umip.org}\footnote{\textit{\href{http://umip.org}{http://umip.org}}}
implementation of NEMO is used, whereas the
\textit{cargeo6.org}\footnote{\textit{\href{http://www.cargeo6.org}{http://www.cargeo6.org}}}
software is used for IPv6 GeoNetworking. ICMPv6 and UDP evaluations in handover
scenarios were performed at INRIA Paris-Rocquencourt campus with the two ARs
previously presented in the testbed description. The speed of the vehicle was
limited to less than 15 km/h, like in a low speed urban scenario. 

The reader can directly click in from~Figure~\ref{fig:handover-ping5-3d-map} to
Figure~\ref{fig:handover-udp2} to see the correspondent results in the AnaVANET
web viewer, to further perceive the details of the gathered results. 

\subsection{ICMP evaluation in a handover scenario}

\begin{figure}[htbp]
   \begin{center}
    \href{http://anavanet.net/demo-vienna/?analysis=1296754401}{
       \includegraphics[width=1\linewidth]{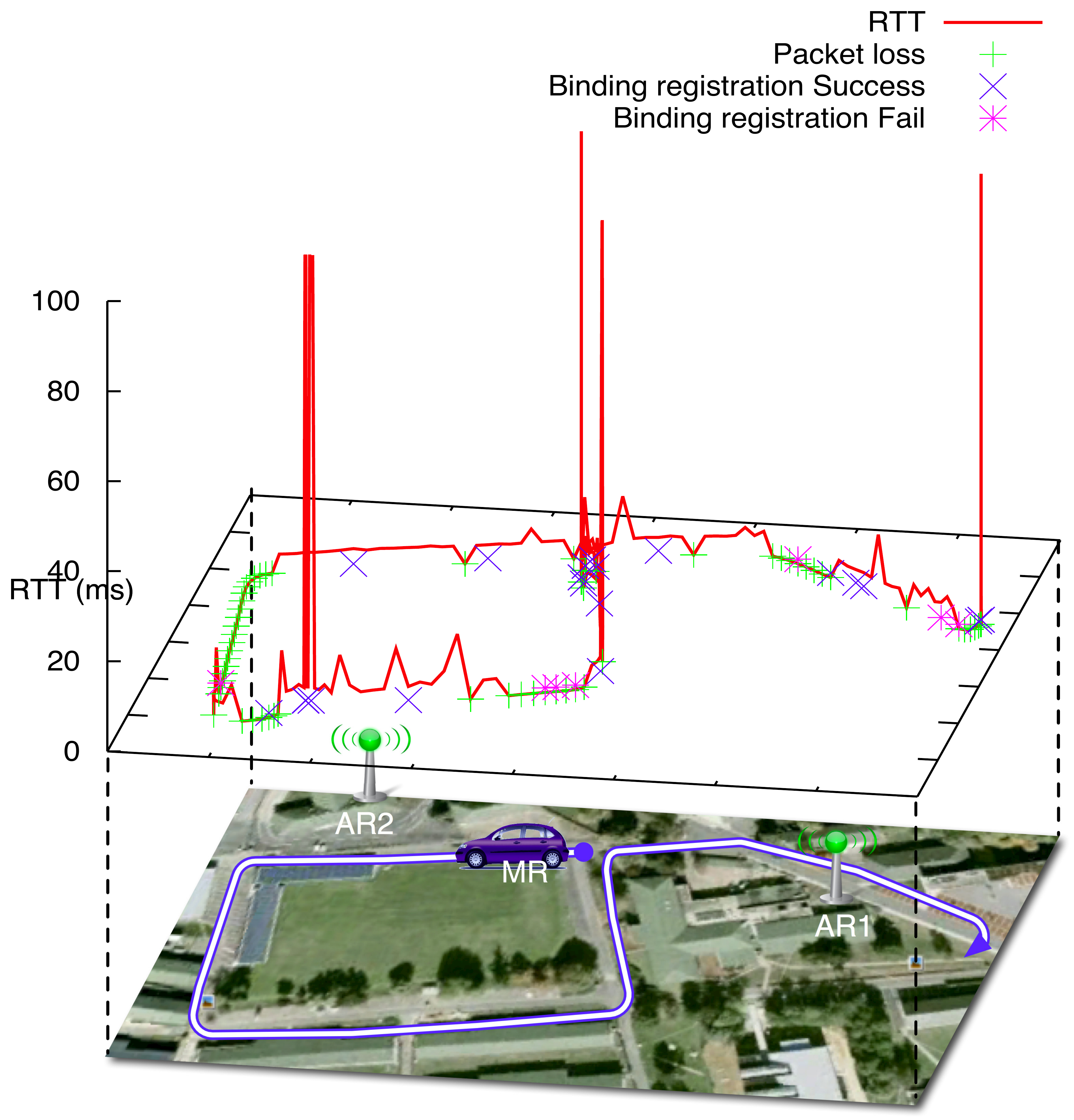}}
      \caption{Map-based RTT, packet losses and mobility signaling in an ICMP evaluation under a handover scenario}
      \label{fig:handover-ping5-3d-map}
    \href{http://anavanet.net/demo-vienna/?analysis=1296754401}{
    \includegraphics[width=1\linewidth]{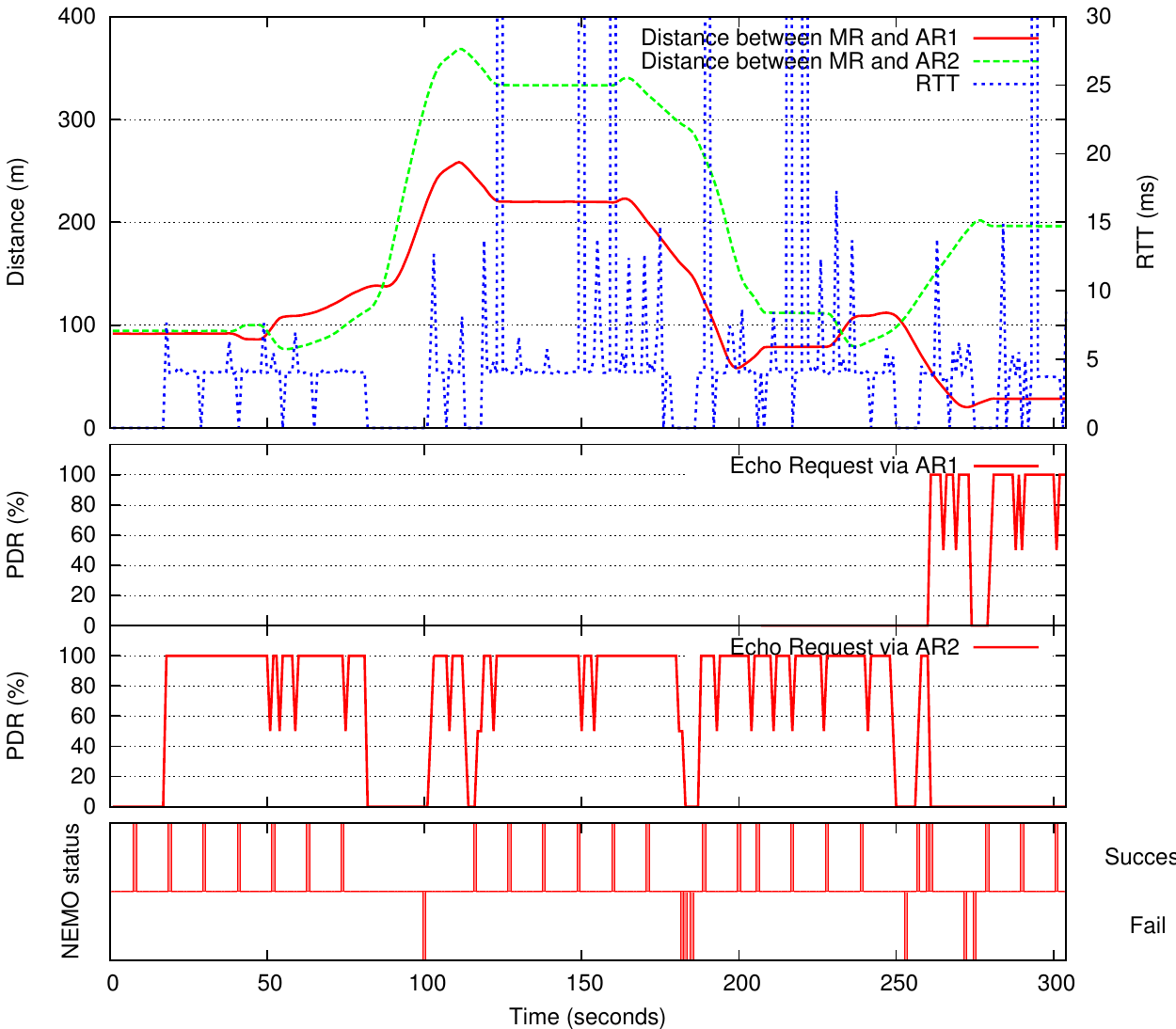}}
      \caption{RTT, packet looses and mobility signaling in an ICMP evaluation under a handover scenario}
      \label{fig:handover-ping5}
  \end{center}
\end{figure}

ICMPv6 echo requests (64 bytes) are sent from the MNN to a common computer
located in the wired network twice a second, which replies with ICMPv6 echo
replies. The results collected in the ICMPv6 tests are plotted in
Figure~\ref{fig:handover-ping5-3d-map}. The lower part shows the itinerary of
the vehicle and the locations of AR1 and AR2 on the map, whereas the upper part
shows the RTT, the packet loss and the result of the mobility signaling. The
X-axis and the Y-axis of the upper part are the latitude/longitude of
the vehicle, corresponding to the road stretch indicated in the lower part of
the figure. When either the request or the reply is lost, the RTT is marked with
a zero value and, at the same time, a packet loss is indicated. A binding
registration success is plotted when the NEMO binding update (BU) and the
corresponding binding acknowledgment (BA) are successfully processed. On the
contrary, if either of them is lost, a binding registration fail is plotted at
the position.

Figure~\ref{fig:handover-ping5} shows the same result of the test, but referred
to the test time. The upper graph shows the RTT and the distance to the two ARs;
the middle one shows the PDR obtained with the two ARs; and, finally, the lower
plot shows the status of the NEMO signaling. A NEMO success means that the
binding registration has been successfully performed, and a fail indicates that
either the BU or the BA has been lost.

As can be seen in Figure~\ref{fig:handover-ping5-3d-map} and
Figure~\ref{fig:handover-ping5}, the RTT is stable at the beginning of the test
near AR2, with a value of around five milliseconds. AR2 is installed at about
100 meters away the road. It sends constant BU messages and, consequently, the
MR successfully performs the binding registration every twelve seconds, without
any packet loss. Soon, after the vehicle turns the first corner (north west of
the square), packets start to be dropped until the second corner. This is
because a near building screens the wireless radio. The binding signaling is
dropped as well in the period. Then it recovers when the vehicle comes to the
straight road on the south. The mobility signaling is successfully sent again
with a regular interval.

The lower straight road of the stretch is less stable than the one on the north,
because of two reasons. First, the location of the south straight road is 250
meters further to AR2 than the one on the north. Thus the signal strength is
weaker now. Second, the trees at this location interfere the wireless radio,
especially at the end of this part of the circuit, as can be seen with the three
consecutive binding registration fails.
When the MR fails to receive a valid matching response within the selected
initial retransmission interval, the MR should retransmit the message until a
response is received. The retransmission by the MR must use an exponential
back-off in which the timeout period is doubled upon each retransmission, until
either the MR receives a response or the timeout period reaches the value of
maximum timeout period as specified in~\cite{rfc6275}. In our particular case
the mobility daemon tries to deliver the BU one second after the first failure
of the binding. Then, when it fails, it increases the retransmission time in
two, four, eight seconds, and so on. 

The performance in the final part of the testing circuit is more stable, and no
binding messages are lost. In this period the vehicle approaches AR2 and then
leaves it turning right at the end of the test.

The MR starts receiving router advertisement (RA) messages from AR1 when the
distance to AR1 is 50 meters, however, the RA messages from AR2 also reaches the
zone. As the result, the vehicle triggers the movement detection, and sends the
mobility signaling via the AR where it receives the RA. When the MR associates
with AR2 some ICMP packets and mobility signaling messages are lost because of
the distance and a near building. When the MR later switches to AR1, the packets
are more stably transmitted.

%

\subsection{UDP evaluation in a handover scenario}

\begin{figure}[htbp]
   \begin{center}
    \href{http://anavanet.net/demo-vienna/?analysis=1296759090}{
    \includegraphics[width=1\linewidth]{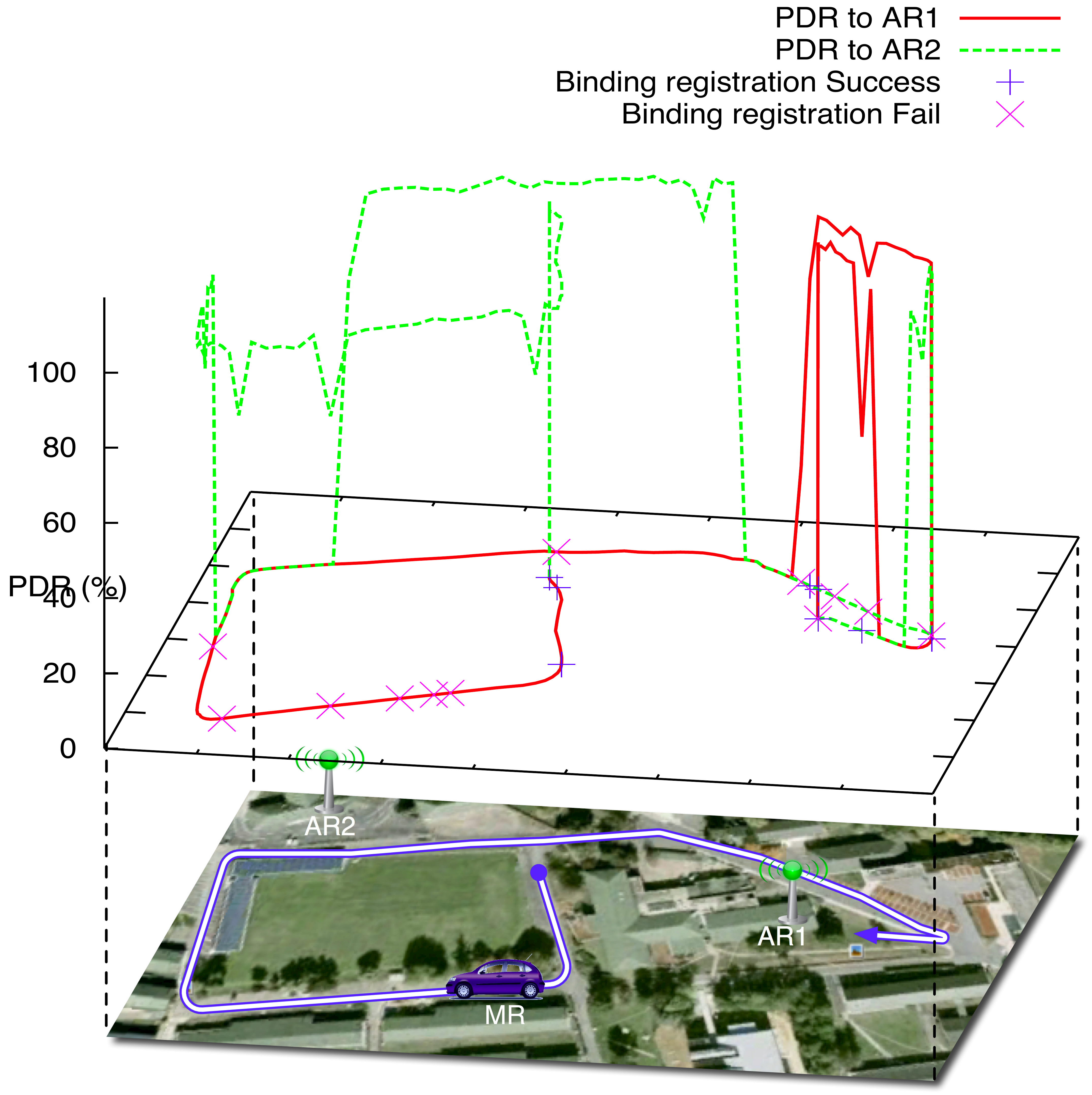}}
      \caption{Map-based PDR of UDP evaluation using NEMO over IPv6 GeoNetworking}
      \label{fig:handover-udp2-3d-map}
    \href{http://anavanet.net/demo-vienna/?analysis=1296759090}{
       \includegraphics[width=1\linewidth]{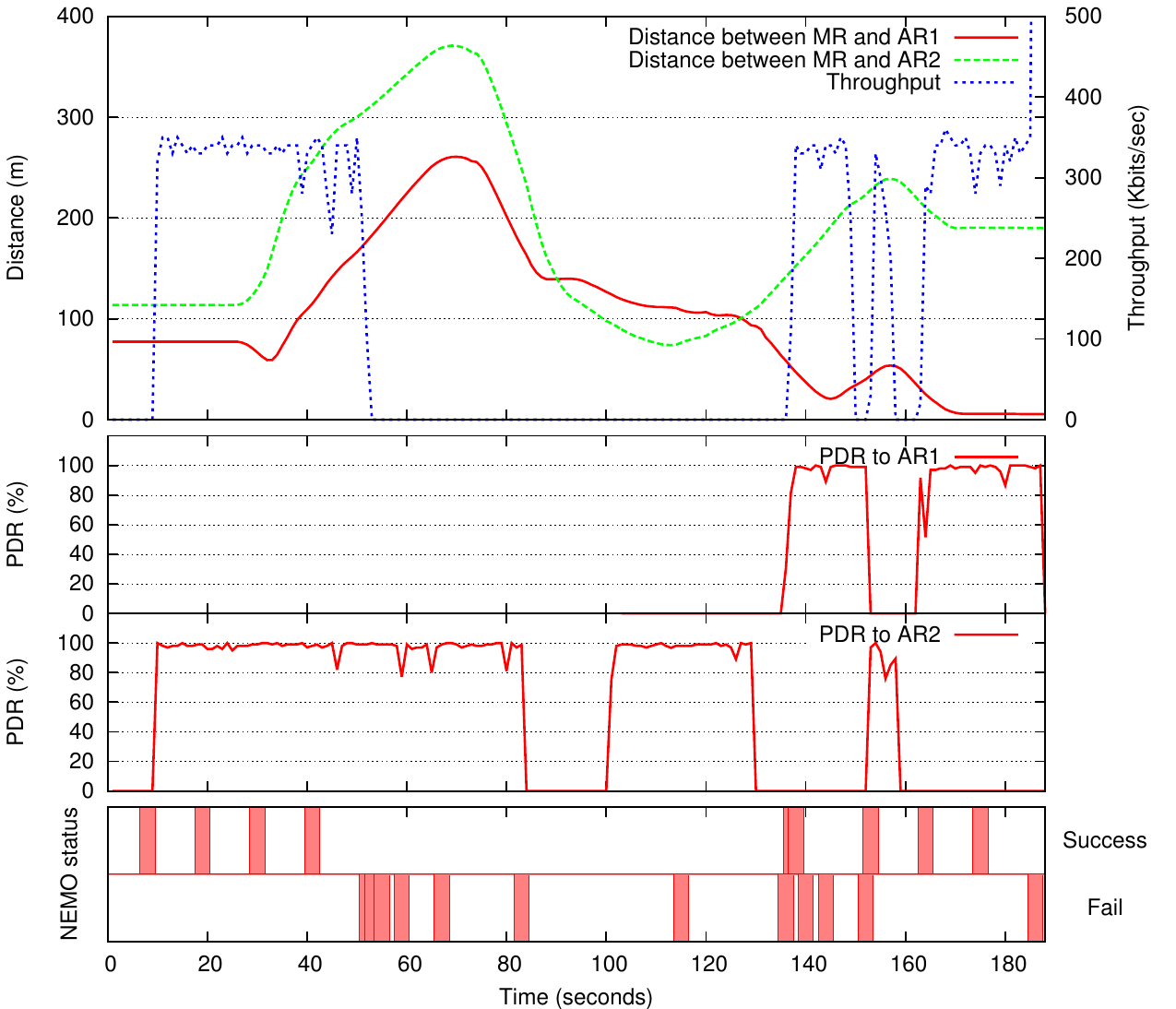}}
      \caption{PDR of UDP evaluation using NEMO over IPv6 GeoNetworking}
      \label{fig:handover-udp2}
  \end{center}
\end{figure}

The results collected in the UDP tests are plotted in
Figure~\ref{fig:handover-udp2-3d-map}. UDP packets are sent from the MNN to the
wired node at a rate of 1 Mbps and a length of 1250 bytes. The lower part of the
figure shows the itinerary of the vehicle, and the upper part corresponds to the
PDR obtained with the ARs and the binding registration results, as in the
previous case. The road stretch is the same one used above, but the vehicle
moves on the contrary direction in this case.

Figure~\ref{fig:handover-udp2} shows the time-mapped results of the same UDP test. The
upper graph shows the UDP throughput from the MNN to the wired node, the middle part
shows the PDR to the two ARs, and the lower plot includes the status of the NEMO
signaling.

The throughput of the UDP traffic is below 30\% of the sending rate of 1Mbps
(\textit{i.e.} 300Kbps), however the PDR with the two ARs reaches 100\%. This is
because the throughput is measured between end nodes (MNN and a node in the
Internet) by \textit{iperf} and the PDR in the wireless links are calculated 
hop-by-hop by AnaVANET. In this case, it shows that more than 70\% of the UDP packets
are dropped outside the wireless links. In fact, the CarGeo6 software experimented a
bottleneck in the processing of so many UDP packets at that time. 
This also explains the phenomenon where the binding registration messages are
lost while none of the UDP packets are lost (this can be seen in the straight
road in the south part of the circuit). In this case, the BUs are lost in the
CarGeo6 software and are not transmitted from the wireless interface.
We can detect where a packet is lost, especially the loss in a wireless
link, thanks to the AnaVANET system (although the cause of the packet losses was
not in the wireless links in the present case). This is because AnaVANET is
capable of measuring both the hop-by-hop network performance and the end-to-end one. 



As can be seen in Figure~\ref{fig:handover-udp2-3d-map}, AR2 is available
most of the test period (especially, around the square) except for the end of
the test. When the vehicle moves in the first straight road in the east, the PDR
to AR2 is almost 100\%. During this period, no binding message is dropped. The
BUs are sent regularly at intervals of twelve seconds. 
%


The packets start being dropped on the west of the square because the building
on the north west corner of the square blocks the wireless radio. When the
beacons exchanged between GeoNetworking nodes twice in a second are dropped, the
correspondent entry of the location table expires in five seconds.

As can be seen in Figure~\ref{fig:handover-udp2}, after the southwest corner, the
end-to-end throughput drops to zero and the binding registration fails, while
the hop-by-hop PDR to AR2 is still almost 100\%. This shows that the mobility
signaling packets are lost in CarGeo6 as explained earlier. Since the binding
life time is configured as 24 seconds, the binding entry in the HA expires 24
seconds after the last successful binding registration. After the expiration of
the binding, HA discards all the packet from the MR. During the period, 
the MR try to send the BUs in exponentially increased interval from 1
second to 32 seconds (1, 2, 4, 8, 16 and 32 seconds). 

Then, at time 139 seconds, when the vehicle is 20 meters away from AR1, the
first binding registration through AR1 successes. UDP packets are switched to
AR1 from this moment. Then at time 155 seconds, the binding registration is
successfully performed via AR2 again. During the handover from AR1 to AR2, from
time 155 seconds to time 158 seconds, three seconds of disconnection are present
in the iperf log. At time 166 seconds, the path to the Internet is switched to
AR1 again. In this handover, UDP packets are lost during four seconds from time
166 seconds.


\section{Qualitative evaluation of the system}\label{lbl:qualitative}

As a result of the experience working with the recent version of AnaVANET, including the results presented above, we have revisited the requirements for an efficient testing environment in vehicular networks detailed in Section~\ref{lbl:issues}, with the aim of evaluating the advantages of the system. Table~\ref{tab:Qualitative-Evaluation} summarizes the most important features, which demonstrate that AnaVANET fulfills the most important requirements and it is an efficient evaluation tool. 

\begin{table}
\begin{centering}
\begin{tabular}{|>{\raggedright}p{2.5cm}|>{\raggedright}p{5.2cm}|}
\hline 
\textbf{\small{}Requirement} & \textbf{\small{}Proposal}\tabularnewline
\hline 
\hline 
{\small{}Path detection} & {\small{}AnaVANET can track the nodes of the communication path for
each transmission}\tabularnewline
\hline 
{\small{}Communication performance in links} & {\small{}The system can measure the PDR of each link as well as the
end-to-end PDR}\tabularnewline
\hline 
{\small{}Geographical awareness} & {\small{}The system outputs the performance indicators in a geo-referenced
way, which facilitates the analysis of results}\tabularnewline
\hline 
{\small{}Intuitive visualization} & {\small{}The movement of vehicles is showed using Google Maps in a
Web application, together with the graphs of the desired performance
metrics. It allows a step-by-step visual analysis of the results.}\tabularnewline
\hline 
{\small{}Independence from network protocols} & {\small{}The system adopts the MAC address for packet tracing. Therefore
any kind of network layer protocol can be evaluated.}\tabularnewline
\hline 
{\small{}Independent from devices} & {\small{}The system does not require specific hardware.}\tabularnewline
\hline 
{\small{}Adaptation to various scenarios} & {\small{}The system can be used in a number of scenarios, including
distance, static, urban and highway tests. Also it allows both V2V
and V2I tests.}\tabularnewline
\hline 
{\small{}Easiness for data collection} & {\small{}The system does not require special software to gather experimental
data. Packet dumps are taken with }\emph{\small{}tcpdump}, {\small{}
and GPS NMEA data is obtained directly from a serial interface, to
finally generate results.}\tabularnewline
\hline 
{\small{}Flexible experimental data format} & {\small{}We have adopted a structured and normalized format defining
a three-layer model in order to increase the flexibility for future
extension.}\tabularnewline
\hline 
\end{tabular}
\par\end{centering}

\protect\caption{Qualitative evaluation of the system\label{tab:Qualitative-Evaluation}}

\end{table}


\section{Conclusions and future work}
\label{lbl:conclusion}

The paper has presented the peculiarities of evaluating vehicular networks experimentally, through
presenting the most used protocols and detailing the needs of the software tools
to be used for this task. After that, the importance of the testing
methodology is described, and a reference design of a vehicular network
evaluation is used to exemplify it. The testbed design and implementation,
testing scenarios, routing protocols and data flows, are found essential to be
fixed beforehand to avoid improvisation during the testing campaign. The
AnaVANET platform is then presented as an efficient evaluation software to
process the data gathered by common testing tools, and then generate lots of performance
indicators of the trials. 
All of these performance parameters are put in the
spatio-temporal context, through the collection and correlation of GPS
information, and most important figures of merit can be exported in the form of
graphics or showed interactively in a web front-end.

The capabilities of AnaVANET are exploited in a novel evaluation of NEMO over
IPv6 GeoNetworking, using the tool to gather RTT, PDR and channel throughput
information. The results reveal that mobile IPv6 connectivity can be maintained
in a V2I case using GeoNetworking over WiFi to pass NEMO IPv6 traffic between
vehicles and infrastructure.

Our future work includes, first, a link layer extension of the system to
analyze the channel quality (RSSI) and load ratio. This data will allow the development of coverage maps for the communication nodes. Second, it is considered the support for multicast data flows, since it is essential for the dissemination of events in vehicular networks. Third, we plan to evaluate a real application developed for 
cooperative ITS.


\section*{Acknowledgment}

This work has been sponsored by the European 7th FP, through the ITSSv6 (contract 270519), FOTsis (contract 270447) and GEN6 (contract 297239) projects, and the Spanish Ministry of Science and Innovation, through the Walkie-Talkie project (TIN2011-27543-C03).

\bibliographystyle{unsrt}
\bibliography{jabref}

\end{document}